\documentclass[10pt]{iopart}
\usepackage{graphicx}
\usepackage{tabularx}
\usepackage{url} 
\usepackage{cite}
\begin{document}

\title{Investigating the fission-like fragments in the $^{12}$C + $^{208}$Pb system at E$^{\star}$ $\approx$ 31.8--45.4 MeV }

\author{Rupinderjeet Kaur$^1$, Amanjot$^1$, Priyanka$^1$, Malika Kaushik$^1$, Subham Kumar$^1$, Arshiya Sood$^2$, Yashraj Jangid$^3$, R. Kumar$^3$ 
, Manoj Kumar Sharma$^4$, Pushpendra P. Singh$^1$ }
%\thanks{deceased}
%\thanks{Presently at: National Institute of Technology Kurukshetra, Kurukshetra - 136 119, Haryana, India}

\address{$^1$Department of Physics, Indian Institute of Technology Ropar, Rupnagar - 140 001, Punjab, India 
$^{2}$ DeSiS team - Institut pluridisciplinaire Hubert Curien-CNRS/Unistra - 67037, Strasbourg, France
$^{3}$Inter-University Accelerator Centre, New Delhi - 110 067, India 
$^{4}$Department of Physics, University of Lucknow, Lucknow - 226 007, Uttar Pradesh, India} 

\ead{pps@iitrpr.ac.in}
\vspace{10pt}
\begin{indented}
\item[\today]
\end{indented}

\begin{abstract}
In this work, the cross-sections of 25 fission-like fragments within the mass range 76$\leq$A$\leq$141, expected to be populated via fission of moderately excited compound nucleus produced as a result of complete and/or incomplete fusion in $^{12}$C+$^{208}$Pb reaction at E$_{\rm lab}$ = 81.9 and 75.8 MeV, have been measured using activation technique followed by offline $\gamma$-ray spectroscopy. The yields of different fission-like fragments have been analyzed to generate isotopic and isobaric yield distributions. The value of the mass dispersion parameter, $\sigma^2_A$, is found to be 2.93 and 2.65 for Antimony (Sb) isotope at excitation energy E$^{\star}$ = 45.4 and 39.6 MeV, and 1.24 for Indium (In) isotope at E$^{\star}$ = 45.4 MeV. The charge dispersion parameter $\sigma_Z$ for Sb is estimated to be 0.769 and 0.714 at E$^{\star}$ = 45.4 and 39.6 MeV, respectively. For In isotopes, the value of $\sigma_Z$ is estimated to be 0.430 at E$^{\star}$ = 45.4 MeV. The value of mass and charge dispersion parameters for Sb and In isotopes have been found to be in good agreement with the values reported in the literature for similar systems. The mass distribution of fission-like fragments is found to be fitted with a Gaussian function, except for a few data points, indicating their population via compound nucleus fission. Further, the mass variance ($\sigma^2_M$) displays linear increment with an increase in excitation energy. Two medically important isotopes, $^{99m}$Tc and $^{111}$In, are populated in this system, suggesting a potential formation route. 
\end{abstract}

%
% Uncomment for keywords
%\vspace{2pc}
%\noindent{\it Keywords}: XXXXXX, YYYYYYYY, ZZZZZZZZZ
%
% Uncomment for Submitted to journal title message
%\submitto{\JPA}
%
% Uncomment if a separate title page is required
%\maketitle
% 
% For two-column output uncomment the next line and choose [10pt] rather than [12pt] in the \documentclass declaration
%\ioptwocol
%

\section{Introduction}
Nuclear fission in heavy-ion (HI) induced reactions has been extensively investigated, generating substantial experimental data across various fission systems. A better understanding of fission dynamics offers crucial insights into various fundamental nuclear processes \cite{bowman1998}. In general, the nuclear reactions involving heavy-mass targets result in the formation of a fully equilibrated compound nucleus (CN), which may decay via light-nuclear particles and/or fission \cite{Bogachev, kozulin2022, tripathi2015, NISHIO201589}. In fission, charge and mass distributions are two crucial experimental observables that are closely linked to the collective dynamics of fission and serve as a valuable testing ground for different theoretical models. An important finding from early investigations was the presence of asymmetric mass distributions in the low-energy fission of most actinides. The asymmetry in the mass distribution was later explained based on nuclear shell effects by incorporating the shell correction term in the liquid drop model \cite{strutinsky1967,wilkins1976}. Exploring the relationship between excitation energy (E*) and fission fragment mass distribution (FFMD) is crucial, as variations in energy strongly influence the resulting mass distribution. The increased excitation energy decreases the mass distribution asymmetry of the fission fragments, leading to symmetric fission. This feature may be explained as a result of a gradual decrease of shell effects with an increase in excitation energies of the compound nucleus as reported in \cite{andreyev2017,chaudhuri2015direct}. Although a combination of the shell and liquid drop model offers insight into the fundamental aspects of nuclear fission, certain aspects still need to be understood. As a result, nuclear fission continues to be an important area of investigation in nuclear physics \cite{SCHMIDT2001169,schmidt2018review,bender2020future}. Extensive research on charge, mass, energy, and angular distribution of fission fragments reported in literature \cite{dubey2016interplay,paul2020measurement,prasad2020systematics,Swinton} has provided more profound insights into the mechanism of fission. Dispersion parameters derived from FFMD provide insights into how the mass distribution varies with changing excitation energy \cite{ghosh2005,pant2001variation}. Dispersion parameters have been obtained for various target-projectile combinations such as  $^{19}$F + $^{169}$Tm \cite{shuaib2019}, $^{11}$B + $^{232}$Th \cite{gubbi1999}, $^{20}$Ne + $^{208}$Pb \cite{tripathi2004}, $^{19}$F + $^{175}$Lu \cite{bhat2021systematic},  $^{12}$C + $^{181}$Ta \cite{kaur2022reaction}, $^{12}$C + $^{197}$Au \cite{kaur2022study}, $^{11}$B + $^{209}$Bi and $^{181}$Ta \cite{karapetyan2016}, $^{20}$Ne + $^{232}$Th \cite{Sodaye} and $^{19}$F + $^{232}$Th \cite{gubbi796}, and their mass variance behavior as a function of excitation energy, mass-asymmetry and angular momentum have also been reported in literature. It is evident from the literature that extensive research has been conducted on fission fragment (FF) dynamics within the actinide region, but the studies on fission fragment (FF) dynamics in the preactinide region are limited. Therefore, we have aimed to contribute valuable insights into the reaction mechanism involving $^{12}$C + $^{208}$Pb system at energies E$_{\rm lab}$ = 81.9 and  75.8 MeV. The entrance-channel properties of $^{12}$C + $^{208}$Pb system, along with the similar systems studied elsewhere, are presented in Table~\ref{table1} for ready reference.

\begin{table*}[t]
\caption{\label{tab:table3}Properties of $^{12}$C + $^{208}$Pb$^{*}$ system with other systems having the same target and different projectile. $\beta$ is the deformation parameter for the projectile and target nuclei, determined by the electric quadrupole transition probability between the 0$^{+}$ ground state and the first 2$^{+}$ state \cite{raman2001}. $\alpha$ is the entrance channel mass-asymmetry and x$_{\rm LD}$ is liquid drop fissility parameter \cite{cohen1974}.}
\centering

\begin{tabular}{cccccccccc}
\hline\hline
Reaction & Barrier  &  \multicolumn{2}{c}{$\beta$ deformation}   & Mass asymmetry  & CN  & x$_{\rm LD}$  & Ref. \\
& height & Projectile& Target & $\alpha$ & & &   \\ 
& (MeV)  & & & & & &   \\  
\hline \\[-2.0ex]

$^{12}$C~+~$^{208}$Pb & 58.58& 0.582 & 0.055 & 0.891 & $^{220 }$Ra  &  0.745  & *  \\

$^{16}$O~+~$^{208}$Pb &76.93 & 0.364 & 0.055& 0.857 &$^{224}$Th   & 0.763  & \cite{itkis1995}  \\               

$^{20}$Ne~+~$^{208}$Pb & 94.93& 0.727& 0.055& 0.824 & $^{228}$U    & 0.781  & \cite{tripathi2004} \\
\hline\hline
\end{tabular}
\label{table1}
\end{table*}

It is important to mention here that Pokrovsky et al. \cite{pokrovsky1999three} reported the fission fragment mass yields (Y), as well as the dependence of the fission fragment total kinetic energy (TKE) and its dispersion $\sigma^2_{TKE}$ on the fission fragment masses of the $^{220}$Ra compound nuclei produced in the $^{12}$C + $^{208}$Pb at  E$_{\rm lab}$ = 57, 59 and 90 MeV. The velocities and coordinates of pair fission fragments were measured using a two-arm time-of-flight spectrometer CORSET. It has been reported that at the two lowest energies, the fission fragment mass distributions are primarily symmetric and can be represented by a Gaussian function for masses between 110 and 125, along with their complementary masses. However, an asymmetric fission mode appears as "shoulders" in the distribution. At an energy of E$_{\rm lab}$ = 90 MeV, the properties of the mass-energy distributions (MEDs) resemble the liquid drop model (LDM) predictions at high temperatures, where Y(M) is Gaussian and TKE(M) forms a parabolic curve. In the present work, the production cross-sections of residues likely to be populated through fission of CN formed via complete fusion and/or incomplete fusion on the $^{12}$C + $^{208}$Pb system have been measured employing the recoil catcher activation technique followed by offline $\gamma$-ray spectroscopy. Data analysis has been done to obtain isotopic yield distribution and mass distribution of fission-like fragments.

 The present paper is organized as follows: Section 2 describes the experimental methodology, Section 3 provides details on data analysis and interpretation of results, and Section 4 presents the summary and conclusions.

\begin{figure}[h]
 \centering
  \includegraphics[height=5cm, width=8cm]{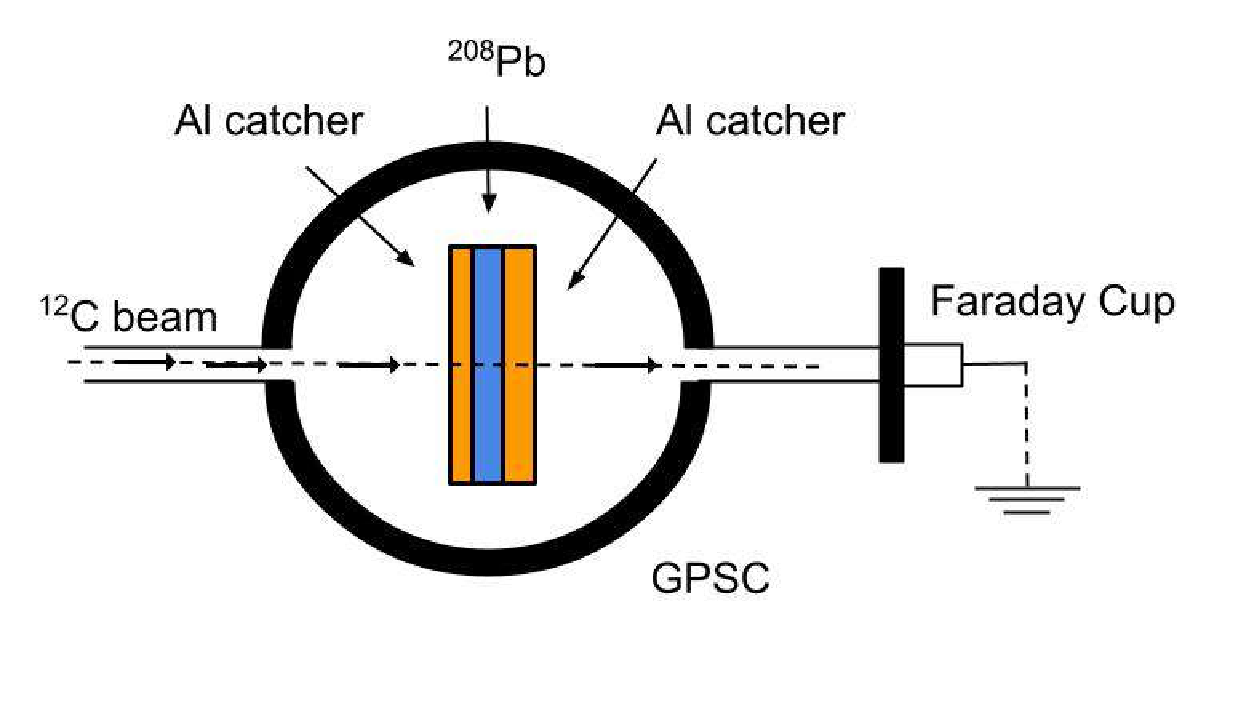}
 \caption{Schematic representation of target-catcher assembly mounted inside the General Purpose Scattering Chamber (GPSC) for irradiation.}
 \label{fig_1}
\end{figure}

\begin{figure}
 \centering
  \includegraphics[height=4.4cm,width=8.5cm]{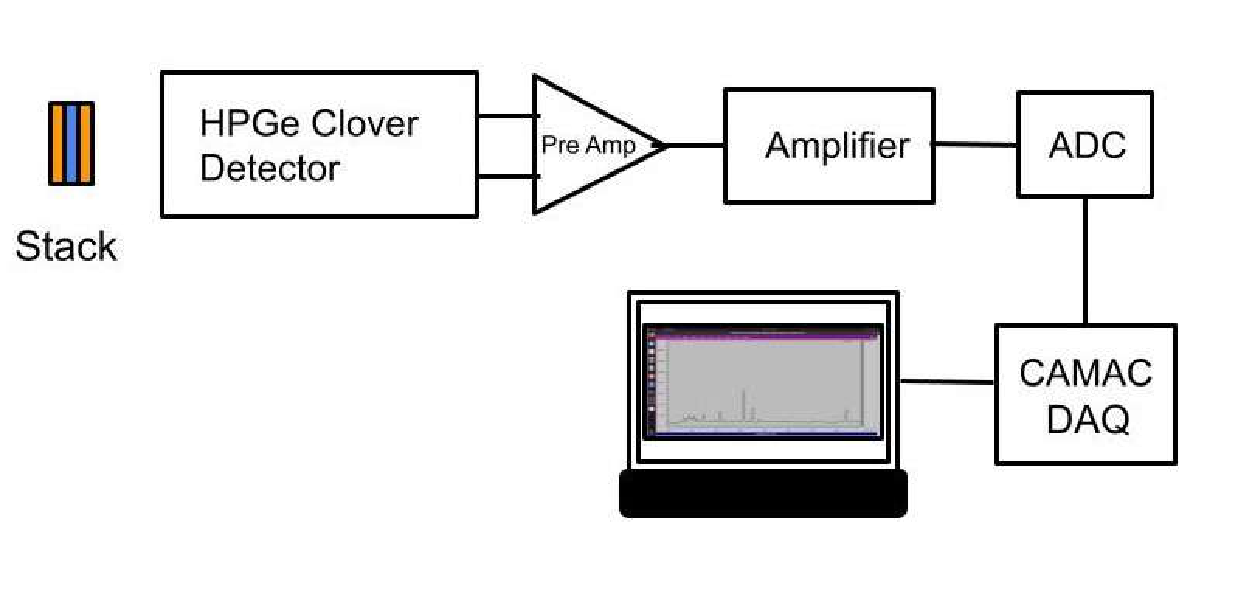}
 \caption{Offline counting setup deploying a clover HPGe detector coupled with a CAMAC-based DAQ system.}
 \label{fig_2}
\end{figure}

\section{Experimental Methodology }
\label{Experimental Methodology}

The experiment was carried out at the Inter-University Accelerator Centre (IUAC), New Delhi, India, in the General Purpose Scattering Chamber (GPSC), equipped with an in-vacuum target transfer facility for swift transfer of irradiated samples from the chamber to the detection setup. Enriched $^{208}$Pb (99.74\%) was utilized to prepare the target foils of thickness $\approx$ 0.169--0.237 mg/cm$^{2}$. $^{208}$Pb were deposited on Al foils (thickness range of $\approx$ 1.0--1.5 mg/cm$^{2}$) using high vacuum evaporation technique \cite{AMANJOT2024113287}. The Al foils were prepared using the cold rolling machine, and their thickness was measured using the $\alpha$ transmission method. Fig.~\ref{fig_1} depicts a stack containing $^{208}$Pb target and Al-catcher foils ($\sim$3.0--4.5 mg/cm$^{2}$), mounted on an aluminum target ladder with a concentric hole of 1.2 cm diameter in GPSC for irradiations. Al-catcher foils were positioned in front and behind the target foil to trap reaction products recoiled out of the target in forward and backward cones.

The irradiations were performed using $^{12}$C beams obtained from 15-UD pelletron accelerator at energies E$_{\rm lab}$ (corrected)~=~81.9 and 75.8 MeV with the beam current ranging between 9--10 nA for $\sim$ 6--9 hours depending upon the half-lives of residues. The beam flux was monitored with a current integrator by measuring the charge collected in the Faraday cup positioned behind the target-catcher assembly. The energy-loss corrected beam energy, i.e., the effective incident beam energy at the half-thickness of the target in each target-catcher foil assembly, was estimated using the code SRIM \cite{srim06}.

After the irradiation, the target-catcher assembly was taken out of the scattering chamber and moved to the offline counting setup. The $\gamma$-ray activities induced in the target-catcher assembly were counted with a pre-calibrated HPGe clover detector coupled to a CAMAC-based DAQ (Data Acquisition) system running on locally developed CANDLE software \cite{ajith2001}. The schematic representation of this setup is shown in Fig.~\ref{fig_2}. For calibration purposes and to estimate the geometry-dependent efficiency of the detectors, standard $\gamma$ sources, i.e., $^{152}$Eu, $^{60}$Co, and $^{133}$Ba, with known strengths were used. The measurements with the standard sources were taken before, in between, and after counting the irradiated target-catcher assembly at different source-detector distances to evaluate detector efficiency. The resolution of the clover detectors was estimated to be $\sim$ 2.3 keV at the 1332 keV $\gamma$-ray of the $^{60}$Co source. The target-catcher assemblies were counted in the same geometry as the standard $\gamma$-ray sources to avoid the solid angle effect during the counting.

\section{Data Analysis}
\begin{figure}
 \centering
  \includegraphics[trim=2.0cm 2.1cm 2.0cm 0.3cm,width=8.4cm]{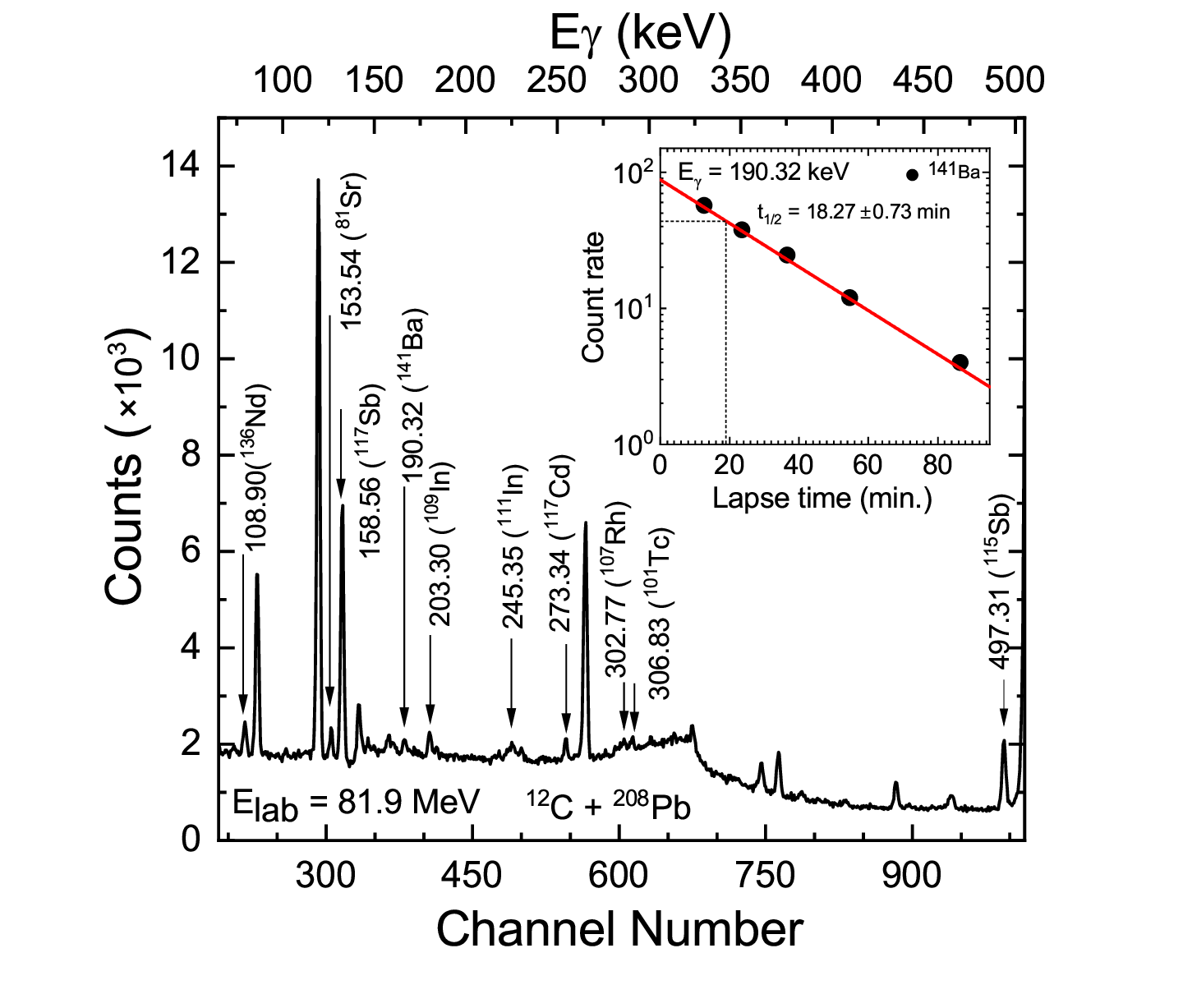}
\vspace{15pt}
     \caption{A portion of $\gamma$-ray spectrum of $^{12}$C+$^{208}$Pb system obtained at E$_{\rm lab}$~=~81.9 MeV. A few reaction products are marked with their decay $\gamma$-lines. Inset shows the decay curve of Barium isotope ($^{141}$Ba) obtained by following 190.32 keV $\gamma$-line at E$_{\rm lab}$~=~81.9 MeV.}
    \label{fig_3}
\end{figure}

The fission-like fragments were identified using the characteristic $\gamma$-rays and their half-lives obtained from the decay curve analysis. The residues of longer half-lives were counted for a week or more. The $\gamma$-ray spectrum of $^{12}$C + $^{208}$Pb system obtained at E$_{\rm lab}$~=~81.9 MeV is presented in Fig.~\ref{fig_3} in which the $\gamma$-lines associated with different fission-like fragments are marked. The inset shows the decay curve of $^{141}$Ba residue obtained following the 190.32 keV $\gamma$-line. As shown in this figure, the activity reduces to half of its value in 18.27 min of lapse time, which shows good agreement with the known characteristic half-live of $^{141}$Ba and confirms its identification. The same procedure has been followed for all the fission-like fragments identified in this work at different energies. In Table~\ref{table2}, the fission-like fragments with their spectroscopic properties, e.g., the half-lives, $\gamma$-ray energies, and intensities taken from the Decay Radiation database within the NuDat (Nuclear Structure and Decay Data) application \cite{nndc} are given.

After identification of all fission-like fragments, their production cross-sections ($\sigma$) were calculated using the standard activation equation given elsewhere \cite{gupta2008}. The error in the production cross-sections may arise due to the uncertainty in target thickness, beam flux, and the geometry-dependent efficiency of the detector. The overall error in the measured cross-sections, including the statistical errors, is estimated to be $<15\%$. In the present work, 25 and 19 fission-like fragments have been identified at E$^{\star}$ = 45.4 MeV (E$_{\rm lab}$ = 81.9 MeV) and E$^{\star}$ = 39.6 MeV (E$_{\rm lab}$ = 75.8 MeV)  respectively. The production cross-sections of fission products, along with the errors, are given in Table~\ref{table2}. Further, an attempt has been made to study the isotopic yield distribution and mass distribution of fission-like fragments in the following sub-sections.

\begin{table}
\caption{Spectroscopic data and production cross-sections of the fission-like fragments identified in $^{12}$C+$^{208}$Pb system.}
\centering

\begin{tabular}{lllllll} 
\hline\hline 
 S. No.  & E$_{\gamma}$ (keV)  & I$_{\gamma}$($\%$)  & t$_{1/2}$ & Nuclide 
   & \multicolumn{2}{c}{$\sigma$ (mb)} \\
   
   & & & &  & 45.4 MeV & 39.6 MeV \\
   
 \hline \\[-2.0ex]
 1 & 406.5 & 12.1 & 14.8 h   & $^{76}$Kr & 10.52$\pm$1.02 &    \\
2 & 589.0 & 39 & 106.3 min    & $^{80}$Sr  & 13.87$\pm$1.27 &  11.65$\pm$1.28  \\ 
3 & 153.54 & 34 & 22.3 min & $^{81}$Sr & 14.50$\pm$1.32 &  12.00$\pm$1.26 \\
4 & 402.58& 50 & 76.3 min   & $^{87}$Kr & 16.80$\pm$1.65 &  14.40$\pm$1.19   \\
5 & 743.36&  93.1 & 16.75 h & $^{97}$Zr & 32.58$\pm$3.37  & 23.07$\pm$2.56 \\
6 & 140.51 & 89 & 6.01 h & $^{99m}$Tc  & 12.54$\pm$1.19 &  7.91$\pm$0.77  \\
7 & 306.83 & 89& 14.2 min  & $^{101}$Tc &  48.34$\pm$4.72 &  33.76$\pm$3.31  \\
 8& 724.21 &  47.8 &4.44 h   & $^{105}$Ru & 49.56$\pm$4.79 & 32.71$\pm$3.25  \\
 9 & 302.77 & 66 & 21.7 min  & $^{107}$Rh & 46.20$\pm$4.50 &  \\
 10 & 203.3 &  74.2 & 4.16 h   & $^{109}$In & 31.79$\pm$2.75 &\\ 
 11 & 657.75 & 98 & 4.92 h & $^{110}$In & 102.49$\pm$10.47 &  65.22$\pm$6.46  \\
 12 & 245.35 &  94.1 &  2.80 d    & $^{111}$In & 50.82$\pm$4.48 &  37.27$\pm$3.52 \\
 13 & 497.31 & 97.9 & 32.1 min   & $^{115}$Sb & 36.95$\pm$3.52  & 20.16$\pm$2.78  \\ 
 14 & 931.84 & 24.8 & 15.8 min  & $^{116}$Sb  & 72.49$\pm$6.66  & 43.28$\pm$5.17  \\ 
15 & 158.56 & 85.9  & 2.80 h  & $^{117}$Sb & 39.72$\pm$3.50 & 22.09$\pm$1.97  \\ 
16 & 273.34 & 27.9 &  2.49 h  & $^{117}$Cd & 55.19$\pm$4.88 & \\
17 & 1050.69 & 97 & 5.00 h  & $^{118m}$Sb  & 5.22$\pm$0.62 & 4.24$\pm$0.52  \\ 
18 & 751.7 & 3.1 & 11.00 min & $^{124}$Ba  & 44.54$\pm$4.07 & 32.50$\pm$3.15 \\
19 & 1114.3 & 38 & 2.10 h & $^{127}$Sn & 37.39$\pm$3.90 & 27.74$\pm$2.74 \\
20 & 536.06 & 99 & 12.36 h  & $^{130}$I & 31.99$\pm$2.94 \\ 
21 & 529.87 & 87 & 20.83 h  & $^{133}$I & 30.29$\pm$2.88 &  \\ 
22 & 847.02 & 96 & 52.5 min  & $^{134}$I & 42.54$\pm$4.01 & 27.62$\pm$2.99\\ 
23  & 249.79 & 90 & 9.14 h   &  $^{135}$Xe & 28.09$\pm$3.01 & 14.33$\pm$1.54 \\
24 & 108.90 & 32 & 50.65 min & $^{136}$Nd & 25.64$\pm$2.41 & 16.31$\pm$1.96 \\
25 & 190.32&  44.8 &  18.27 min   & $^{141}$Ba   & 15.04$\pm$1.57  & 12.98$\pm$1.68 \\

\hline\hline
\end{tabular}
\label{table2}
\end{table}

\subsection{ Isotopic yield distribution}
In a heavy composite system, the emission of particles (charged and/or uncharged) competes with the process of fission, particularly at moderate excitation energies. Therefore, the emission of particles from the fission fragments may give rise to the isotopic and isobaric distributions of the fission residues. However, the emission of charged particles faces a significant challenge due to the presence of the Coulomb barrier. The fact that neutron emission is more probable than proton emission, results in the observation of isotopic yield distribution in most cases. The study of isotopic yield distribution insights into the major characteristics of low energy fission \cite{donzaud1998low}. 
\begin{figure} 
 \centering
    \includegraphics [trim=2.0cm 0.7cm 2.0cm 0.7cm,width=20cm]{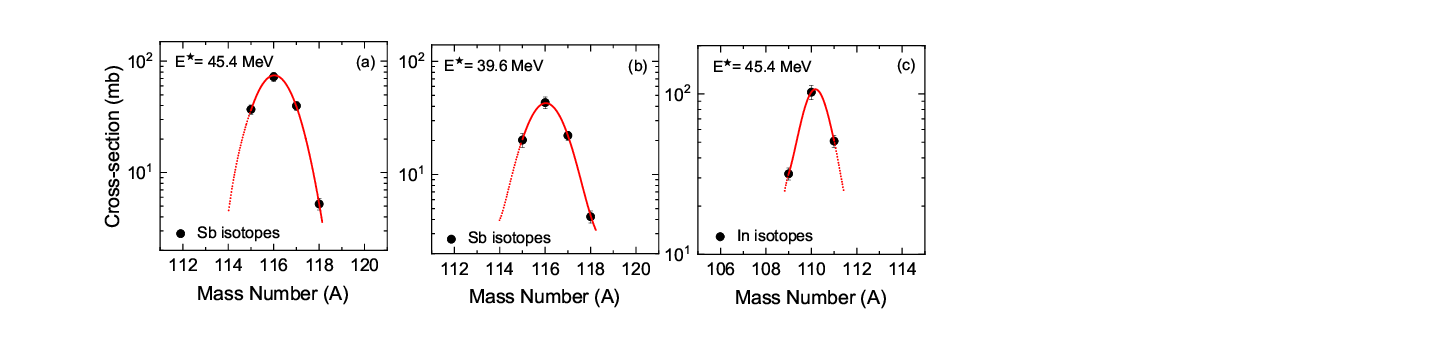}
 \caption{Isotopic yield distribution of antimony $^{115,116, 117,118m}$Sb at E$^{\star}$ = (a) 45.4 MeV, (b) 39.6 MeV, and indium isotopes $^{109,110,111}$In at E$^{\star}$ = (c) 45.4 MeV. The solid line is the Gaussian fit to the experimental data, and the dotted line represents an extension of the fitting function to achieve the variance parameter.}
    \label{fig_4}
\end{figure}

In the present work, four isotopes of antimony ($^{115,116,117,118m}$Sb) and three isotopes of indium ($^{109,110,111}$In) have been identified at E$^{\star}$ = 45.4 and 39.6 MeV, and at E$^{\star}$ = 45.4 MeV, respectively. The experimentally measured yields of Sb and In isotopes have been analyzed as per the prescription presented in Ref. \cite{sharma2011} to achieve isotopic yield distribution and to obtain the charge distribution parameters. The parameters of isotopic yield distribution, e.g., most probable mass $A$$_{p}$ and width parameter $\sigma_{A}$, for antimony and indium isotopes at E$^{\star}$ = 45.4 and 39.6 MeV were obtained by fitting a Gaussian function to their respective yields as shown in Figs.~\ref{fig_4}(a)-~\ref{fig_4}(c), respectively. The values of most probable mass $A$$_{p}$ for antimony isotopes at excitation energy E$^\star$ = 45.4 and 39.6 MeV estimated to be 116.027 $\pm$ 0.014 and 116.029 $\pm$ 0.008 MeV, and the width parameters for isotopic yield distribution are estimated to be 1.711 $\pm$ 0.012 and 1.628 $\pm$ 0.005, respectively. For indium isotopes, the values of $A$$_{p}$ and $\sigma_{A}$ at E$^\star$ = 45.4 MeV is found to be 110.178 $\pm$ 0.005 and 1.113 $\pm$ 0.008. The uncertainties in these estimated parameters are the fitting errors.

\begin{table}
\caption{Comparison of variance parameter ($\sigma^2_{A}$) of isotopic yield distributions populated in different projectile-target combinations.}
\centering
\begin{tabular}{lllll} 
\hline\hline 
 System  &  E$^\star$ (MeV)  & Element  & $\sigma^2_{A}$  & Ref.\\ [0.5ex] 
 \hline \\[-2.0ex]

 $^{12}$C~+~$^{208}$Pb & 45.4 & Sb   & 2.93   & a \\

 $^{12}$C~+~$^{208}$Pb &  39.6 & Sb   & 2.65   & a \\

 $^{12}$C~+~$^{208}$Pb & 45.4 & In   & 1.24   & a \\

 $^{20}$Ne~+~$^{208}$Pb & 46.4& Sb  & 3.43$\pm1.02$   & \cite{tripathi2004} \\

$^{20}$Ne~+~$^{208}$Pb & 46.4& I  & 3.95$\pm0.87$   & \cite{tripathi2004} \\

$^{7}$Li~+~$^{232}$Th & 41.7 & Sb   & 4.08   & \cite{tripathi2002} \\

 $^{7}$Li~+~$^{232}$Th & 41.7 & I   & 3.96  & \cite{tripathi2002}\\

 $^{11}$B~+~$^{232}$Th & 55.7 & Sb   & 4.00   & \cite{gubbi1999} \\

 $^{11}$B~+~$^{232}$Th & 55.7 & I  & 5.43   & \cite{gubbi1999} \\

 $^{11}$B~+~$^{238}$U & 67.4& Rb  & 3.84$\pm0.16$   & \cite{de1976} \\

 $^{11}$B~+~$^{238}$U & 67.4& Cs  & 3.95$\pm0.14$   &  \cite{de1976} \\

 $^{22}$Ne~+~$^{238}$U & 64.5& Rb  & 4.23$\pm0.40$   &   \cite{de1976}\\

 $^{22}$Ne~+~$^{238}$U & 64.5& Cs  & 4.26$\pm0.90$   & \cite{de1976} \\

 $^{16}$O~+~$^{169}$Tm & 61.0 & Tc  & 4.62 & \cite{ppsingh2008}  \\  

$^{16}$O~+~$^{169}$Tm & 61.0 & In  & 4.24 &   \cite{ppsingh2008}  \\

 $^{12}$C~+~$^{169}$Tm & 68.6 & Kr  & 3.90$\pm$0.20 & \cite{sood2017} 
 \\  
$^{12}$C~+~$^{169}$Tm & 68.6 & Tc  & 3.27$\pm$0.18 &   \cite{sood2017}  \\

 $^{19}$F~+~$^{169}$Tm & 69.4 & Nd  & 4.92$\pm$0.80 &  \cite{shuaib2019} \\  

$^{19}$F~+~$^{169}$Tm & 69.4 & In  & 4.49$\pm$1.10 &   \cite{shuaib2019}  \\
\hline\hline
$^{a}$Present work

\end{tabular}
\label{table3}
\end{table}

\begin{figure} 
\centering
    \includegraphics[trim=1.8cm 1.0cm 2.0cm 2.0cm,width=8cm]{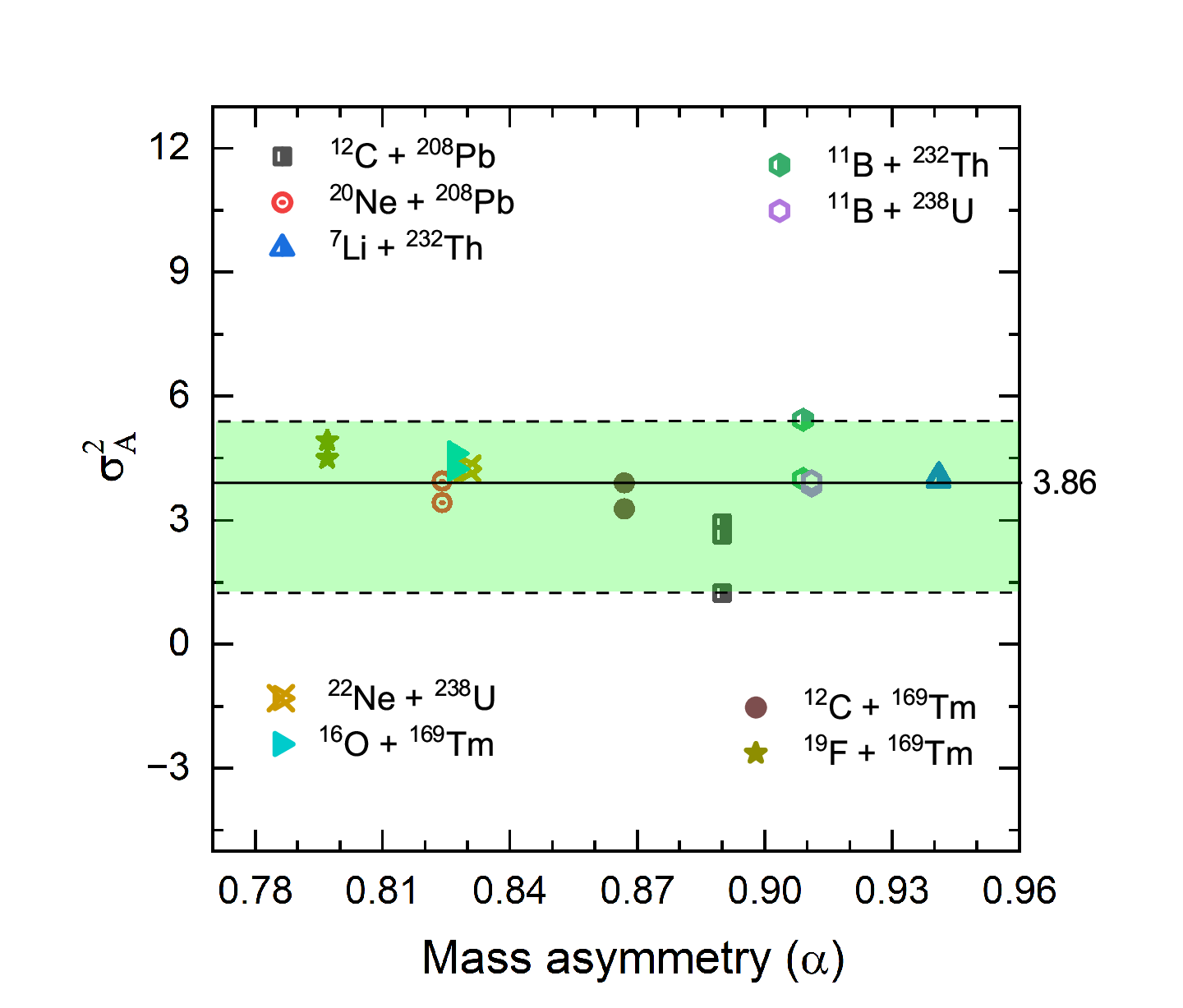}
\vspace{5pt}
    \caption{The value of variance parameter ($\sigma^2_A$) obtained from the analysis of isotopic yield distributions plotted with mass-asymmetry. The spread in $\sigma^2_A$ for different systems is shown within the dashed horizontal lines. The solid line at $\sigma^2_A$ = 3.86 represents the average value of the variance parameter in the presented systems. }
    \label{fig_5}
\end{figure}

\begin{table}
  \caption{The value of variance parameter ($\sigma^2_{A}$) of different isotopes obtained from the isotopic yield distribution and mass-asymmetry parameter ($\alpha$) of several fissioning systems.}
  \centering

\begin{tabular}{lllll}
\hline\hline
 Target &  Projectile   & Element & $\alpha$ & $\sigma^2_{A}$  \\ [0.5ex] 
 \hline & \\[-2.0ex]
$^{208}$Pb & $^{20}$Ne & Sb   & 0.82 & 3.43 \\

$^{208}$Pb & $^{12}$C  & Sb   & 0.89   & 2.92 \\
 
$^{238}$U & $^{22}$Ne & Rb   & 0.83   & 4.23\\

$^{238}$U & $^{11}$B & Rb  & 0.91  & 3.84\\

$^{238}$U & $^{22}$Ne & Cs   & 0.83   & 4.26\\

$^{238}$U & $^{11}$B & Cs  & 0.91   & 3.95\\

$^{232}$Th & $^{11}$B & I  & 0.91   & 5.43\\

$^{232}$Th & $^{7}$Li & I & 0.94   & 3.96\\

$^{169}$Tm & $^{16}$O & Tc  & 0.83  &4.62 \\

$^{169}$Tm & $^{12}$C  & Tc  & 0.87   &3.27\\

$^{169}$Tm & $^{19}$F & In  & 0.80   &4.49 \\

$^{169}$Tm & $^{16}$O & In  & 0.83   &4.24 \\
\hline\hline

\end{tabular}
\label{table4}
\end{table} 

The variance parameter ($\sigma^2_A$) for Sb isotopes at excitation energy E$^\star$ = 45.4 and 39.6 MeV is found to be 2.93 and 2.65, and for In isotopes is found to be 1.24 at E$^\star$ = 45.4 MeV, respectively. The value of $\sigma^2_{A}$ for several other systems at comparable excitation energies is shown in  Table~\ref{table3} along with the present system. As can be noticed from this table, the value of $\sigma^2_{A}$ for the present system agrees reasonably well with those reported in the literature and give confidence to the present observations. For better insights into the spread of isotopes in different systems (given in Table~\ref{table3}), the value of variance parameter ($\sigma^2_{A}$) of different isotopes and the mass-asymmetry parameter ($\alpha$) for various projectile-target combinations are plotted in Fig.~\ref{fig_5}. As shown in this figure, the value of variance for these isotopes falls within a range of width 4.13. The average value of the variance parameter is found to be 3.86, suggesting the involvement of fusion-fission dynamics in the population of Sb and In isotopes. It is in line with the observations reported in the literature. The data presented in Table~\ref{table4} suggests a higher value of variance parameter ($\sigma^2_A$) for less mass-asymmetric systems for the same target.

 \subsection{ Isobaric yield distribution}
\begin{figure} 
 \centering
    \includegraphics[trim=2.0cm 0.7cm 2.0cm 0.7cm,width=20cm]{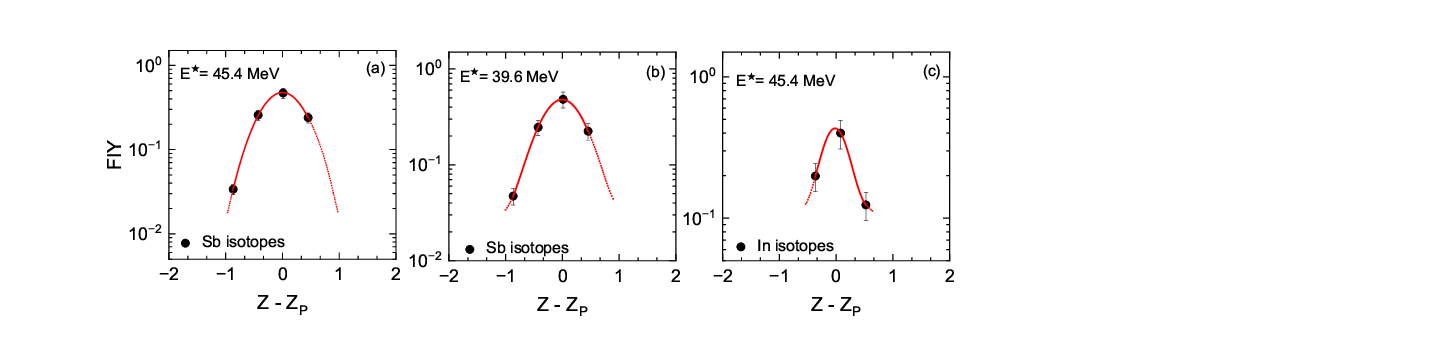}
 \caption{Fractional yield (FY) of Sb isotopes ($^{115,116,117,118m}$Sb) as a function of $Z$ - $Z$$_p$ at E$^\star$  = (a) 45.4 MeV, (b) 39.6 MeV and indium isotopes ($^{109,110,111}$In) at E$^\star$ = (c) 45.4 MeV. The lines and symbols are self-explanatory.}
    \label{fig_6}
\end{figure}

The isobaric charge distribution is a crucial post-fission observable. As demonstrated by Gubbi et al. \cite{gubbi1999}, to deduce the total chain yield of a given fission residue of mass A, it is essential to have the information on the isobaric charge dispersion parameter ($\sigma_{Z}$) and the most probable charge ($Z$$_p$) for the fission product with the highest yield among all the products of a given mass chain $A$. The $Z$$_p$ for the antimony and indium isotopes are calculated using the relation:    
 
\begin{equation}
Z_p(A) = \frac{Z} {A_p}A , 
\end{equation}

where $Z$ and $A$ represent the charge and mass of the fission-like fragments, respectively, while $A$$_{p}$ signifies the most probable mass. In this study, an effort has been made to analyze the isobaric charge distribution in terms of fractional independent yield (FIY). The fractional independent yields for isotopes Sb and In were obtained by dividing the independent yields by their corresponding charge yields. The distribution of fractional chain yield vs charge corrected isotopic fragments ($Z$ - $Z$$_p$) is shown in Figs.~\ref{fig_6}(a)-~\ref{fig_6}(c), respectively. The charge dispersion parameter ($\sigma_{Z}$) obtained from the Gaussian fit to the data for Sb is estimated to be 0.769 and 0.714 at E$^{\star}$ = 45.4 and 39.6 MeV, respectively. For In isotopes, the value of $\sigma_{z}$ is estimated to be 0.430 at E$^{\star}$ = 45.4 MeV.

\begin{table}
\caption{The isobaric charge dispersion parameter obtained from the analysis of data presented in Figs.~\ref{fig_6}(a)-~\ref{fig_6}(c) at different excitation energies (E$^\star$).}
\centering
\begin{tabular}{cccc}
\hline\hline

E$^\star$ & Element &   $\sigma$$_{Z}$ & $\sigma$$_{Z}$  \\[0.5ex] 

(MeV) & &  From  Figs.~\ref{fig_6}(a)-~\ref{fig_6}(c) &  From Eq.(2)\\ [0.5ex] 

 \hline \\[-2.0ex]
   45.4  &Sb  & 0.769  & 0.752   \\
  39.6  &Sb  & 0.714  & 0.715    \\
  45.4  &In  & 0.430  & 0.495   \\
  \hline\hline
 \end{tabular}
\label{table5}
\end{table}

Additionally, the charge dispersion parameter ($\sigma_{Z}$) has been calculated by converting the width parameter ($\sigma_{A}$) of isotopic yield distribution into $\sigma_{Z}$ using the formulation given below,

\begin{equation}
\sigma_Z = \frac{\sigma_A} {A_p} Z.
\end{equation}

The values of $\sigma_{Z}$ obtained for different isotopes as a function of energy are reported in Table~\ref{table5}. As can be seen from this table, the values of $\sigma_{Z}$ obtained from the analysis of charge distribution data presented in Figs.~\ref{fig_6}(a)-~\ref{fig_6}(c) and the one calculated from the above formulations, are found to be in good agreement, indicating the self-consistency of the approaches used in the present analysis.

\subsection{Mass distribution of fission-like fragments}
\begin{figure}
 \centering
    \includegraphics[trim =2.5cm 0.3cm 2.5cm 0.7cm,width=11.0cm]{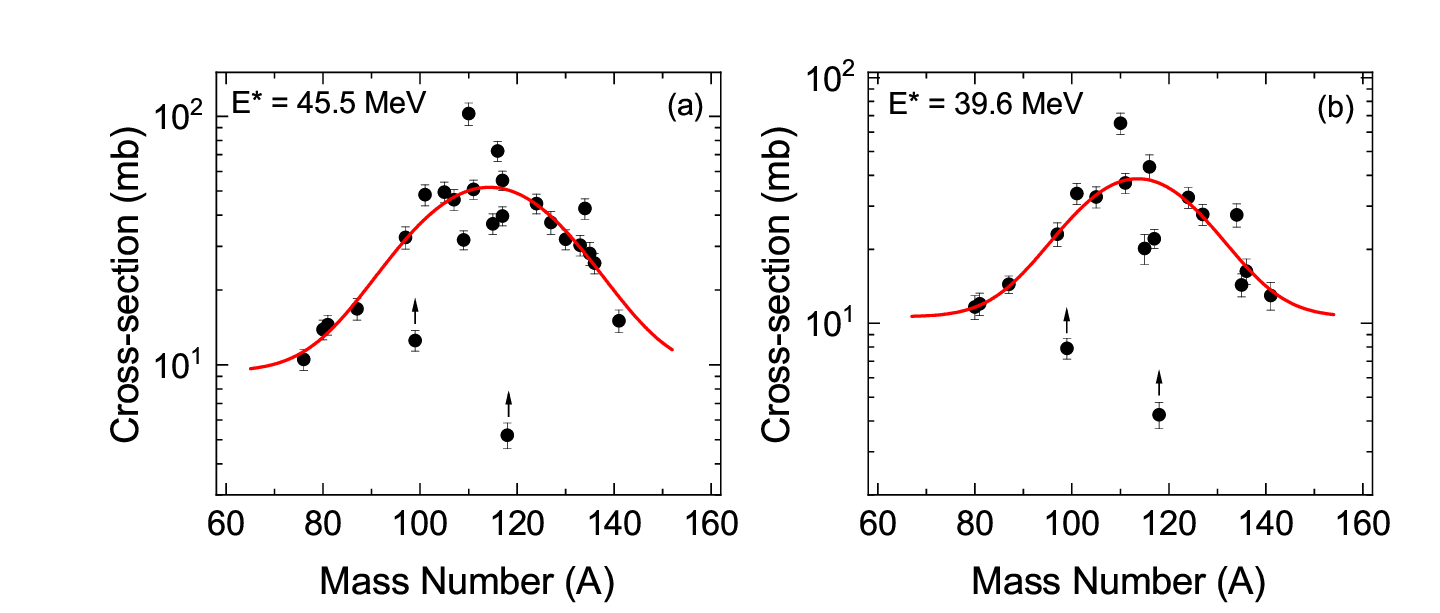}
  \caption{Mass distribution of fission fragments in the $^{12}$C+$^{208}$Pb system at excitation energies E$^{\star}$ = (a) 45.4 MeV and (b) 39.6 MeV. The solid lines show the Gaussian fit through data points. } 
    \label{fig_7}
\end{figure}

The mass distribution of fission fragments is a crucial post-fission observable directly linked to the collective dynamics of the fission \cite{wagemans1991}. The activities produced in the target-catcher assembly were used to determine the mass distribution. To generate mass distributions, the experimentally determined production cross-sections of fission-like fragments produced in the $^{12}$C + $^{208}$Pb system are plotted as a function of mass number (A) in Figs.~\ref{fig_7}(a)-~\ref{fig_7}(b) at E* = 45.4 and 39.6 MeV respectively. The upward arrows indicate that only the metastable states have been measured, and the total production cross-sections of these fission-like fragments are expected to increase. From these figures, it can be seen that the observed mass distribution is Gaussian-like and fitted with a Gaussian function suggesting the formation of these fission-like fragments from the compound nuclear process. The solid lines represent the Gaussian fit through data points. The present observation is in line with the observations reported by  Pokrovsky et al. \cite{pokrovsky1999three}. The centroid (M$_p$) and width ($\sigma_M$) of the mass distribution curve at E* = 45.4 MeV are found to be 114.29 $\pm$ 0.50 and 15.38 $\pm$ 0.82 respectively. At E* = 39.6 MeV, the value of M$_p$ and $\sigma_M$ are found to be 113.56 $\pm$ 0.32 and 12.99 $\pm$ 0.56 respectively. The quoted uncertainties are the fitting errors. The value of full width at half maximum (FWHM) at E* = 45.4 and 39.6 MeV is found to be 36.22 and 30.61 respectively. The FWHM of the distributions is comparable with those observed in $^{20}$Ne + $^{209}$Bi \cite{hinde1989systematics}, $^{20}$Ne + $^{208}$Pb \cite{tripathi2004}, $^{16}$O + $^{208}$Pb \cite{itkis1995}. The fission dynamics are significantly impacted by the predominant direction of mass flow within the dinuclear system. As reported in the literature \cite{ramamurthy1990, kailas2007, prasad2010}, if the mass asymmetry ($\alpha$) is greater than the Businaro-Gallone critical mass asymmetry ($\alpha$$_{\rm BG}$) \cite{abe1986kek}, mass flows from the projectile to the target, leading to the formation of the compound nucleus which may eventually decay via the process of fission. If the value of $\alpha$ $<$ $\alpha_{\rm BG}$, the mass flow occurs from the target to the projectile, and a dinuclear system is formed, decaying before equilibrating in all degrees of freedom. For $^{12}$C + $^{208}$Pb system, the critical mass asymmetry ($\alpha_{\rm BG}$) is 0.86. In contrast, the mass asymmetry ($\alpha$) is 0.89, i.e., $\alpha$ $>$ $\alpha_{\rm BG}$, resulting in the occurrence of fission via compound nucleus formation path. 

\begin{figure} 
\centering
    \includegraphics[trim=1.8cm 1.0cm 2.0cm 2.0cm,width=8cm]{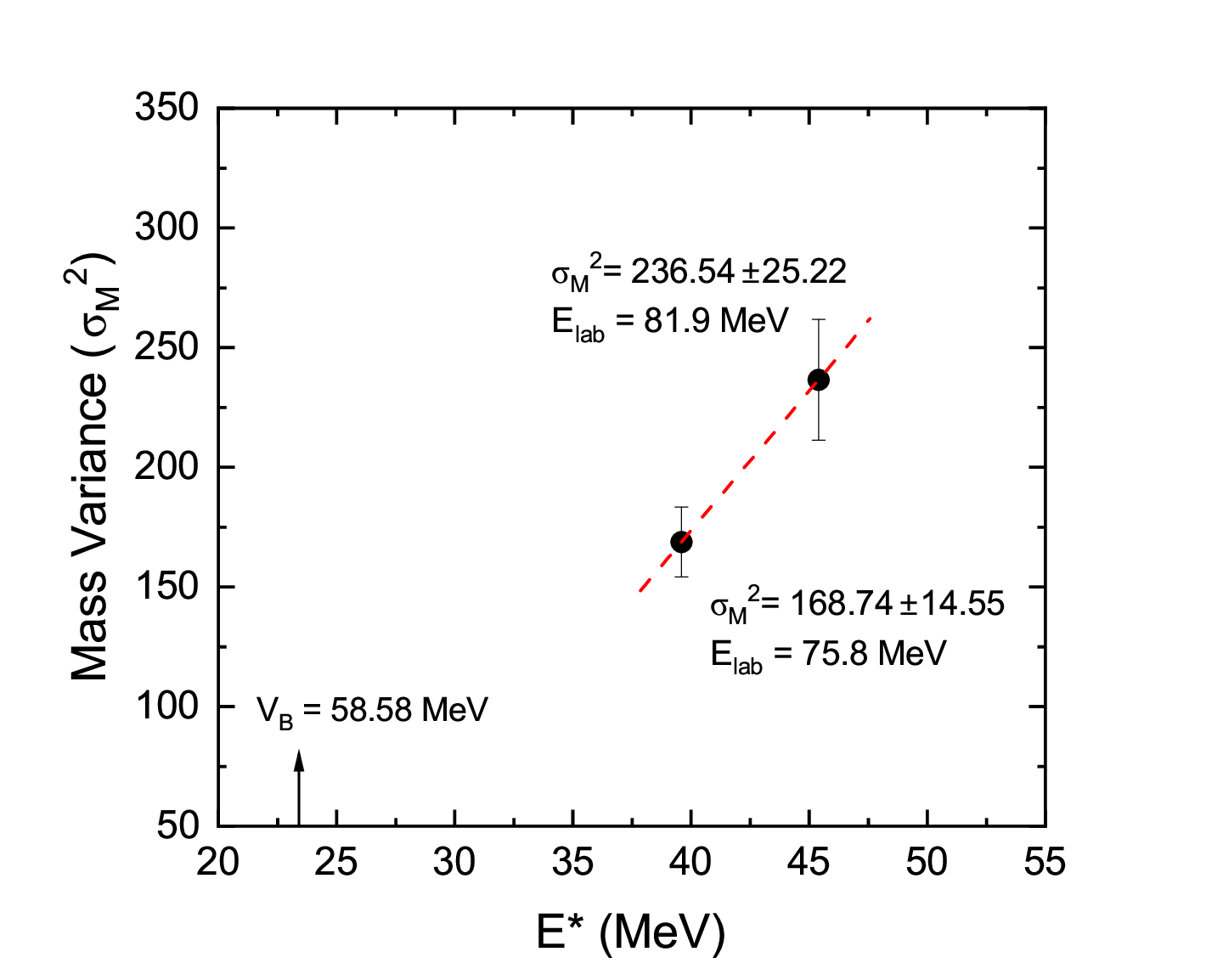}
    \caption{Mass variance ($\sigma^2_M$) as a function of excitation energy (E$^{\star}$) for $^{12}$C+$^{208}$Pb system. The dotted line shows the linear increase in $\sigma^2_M$ with E$^{\star}$. The vertical arrow indicates the excitation energy corresponding to the Coulomb barrier of the system. }
    \label{fig_8}
\end{figure}

 In order to study the change in mass variance ($\sigma^2_{M}$) with excitation energy, the value of $\sigma^2_{M}$ is plotted as a function of excitation energy in Fig.~\ref{fig_8}. The value of $\sigma^2_{M}$ increases with excitation energy, indicating a large spread in the fission fragment masses at higher excitation energies. The observed variation in the value of $\sigma^2_{M}$ with excitation energy above the barrier follows the same trend as reported by Ghosh et al. \cite{ghosh2005}. It is important to mention that this study only examined the system at two energies above the barrier. Consequently, additional research is required to explore how the value of  $\sigma_M^{2}$ changes with excitation energy at and below the barrier energies for a more comprehensive understanding.

\section{Summary and conclusions}
The production cross-sections of various fission-like fragments populated in the $^{12}$C + $^{208}$Pb system have been measured at E$^{\star}$ = 45.4 and 39.6 MeV. The isotopic and isobaric yield distributions of two fission-like fragments antimony (Sb) and indium (In)  have been obtained. It has been found that a single Gaussian function effectively explains the isotopic yield distributions for Sb and In isotopes. The derived mass and charge dispersion parameters agree reasonably well with the experimental values reported in the literature for other fissioning systems. The values of variance for many such distributions in literature and obtained in the present study for Sb and In fall within a narrow range of width 4.13. The data analysis further indicates that the entrance-channel mass asymmetry significantly affects the variance of isotopic yield distribution. The isobaric distribution has been obtained for Sb and In isotopes from the analysis of their experimental yields to get the charge dispersion parameter $\sigma_{Z}$. The value of $\sigma_{Z}$ for Sb was found to be 0.769 and 0.714 at E$^{\star}$ = 45.4 and 39.6 MeV, respectively. For In isotopes, $\sigma_{Z}$ was found to be 0.430 at E$^{\star}$ = 45.4 MeV. These values of $\sigma_{Z}$ show a good agreement with the values of $\sigma_{Z}$ calculated by substituting the obtained isotopic distribution parameter in theoretical expression. The agreement between these values shows the self-consistency of the present analysis and the experimental data. The mass distribution of fission-like fragments has been studied at different excitation energies to investigate the dispersion of fission-like fragments. The mass distributions at two studied energies are found to be symmetric and fitted with a Gaussian function, indicating the population of these fission-like fragments via deexcitation of the compound nucleus. It has been observed that the mass variance increases with increases in excitation energy. In the present study, medically important $^{99m}$Tc and $^{111}$In isotopes with decent production cross-section have been observed. As such, the fission of excited $^{220}$Ra nuclei may provide an alternative path to producing $^{99m}$Tc and $^{111}$In isotopes for application in radio medicine. For better insights into the fission dynamics of $^{220}$Ra, an online experiment to measure the neutron multiplicity is in order.

\section*{Acknowledgements}
The authors acknowledge the Inter-University Accelerator Centre, New Delhi, for the necessary facilities to perform these measurements, the Pelletron crew for hassle-free delivery of $^{12}$C beams throughout the run, and the Indian Institute of Technology Ropar for a grant to procure the enriched target material for this experiment. One of the authors, RK, thanks the University Grant Commission (UGC) in New Delhi, India, for a doctoral fellowship. Another author, MKS, thanks the DST, Delhi, India, for the financial support under the SERB-SURE grant.

\section*{References}

\bibliographystyle{iopart-num}
\bibliography{iopart-num}

\end{document}